\documentclass{ws-mpla}

\begin{document}

\markboth{BO-QIANG MA} {Muon Charge Ratio of Ultrahigh Energy Cosmic
Rays}

%%%%%%%%%%%%%%%%%%%%% Publisher's Area please ignore %%%%%%%%%%%%%%
\catchline{}{}{}{}{}
%%%%%%%%%%%%%%%%%%%%%%%%%%%%%%%%%%%%%%%%%%%%%%%%%%%%%%%%%%%%%%%%%%%

\title{Muon Charge Ratio of Ultrahigh Energy Cosmic Rays\footnote{Invited talk at 2007 International Symposium on Cosmology
and Particle Astrophysics (CosPA2007), November 13-15, 2007,
Taipei.} }

\author{\footnotesize BO-QIANG MA}

\address{School of Physics and State Key Laboratory of Nuclear Physics and
Technology, Peking~University, Beijing 100871, China\\
mabq@phy.pku.edu.cn}

\maketitle

%\pub{Received (Day Month Year)}{Revised (Day Month Year)}

\begin{abstract}
The muon charge ratio of ultrahigh energy cosmic rays may provide
information to detect the composition of the primary cosmic rays. We
propose to extract the charge information of high energy muons in
very inclined extensive air showers by analyzing their relative
lateral positions in the shower transverse plane.

%\keywords{Keyword1; keyword2; keyword3.}
\end{abstract}

%\ccode{PACS Nos.: include PACS Nos.}

\vspace{6mm}

The most high energy particles can be observed by human being are
from cosmic rays. The study of them belongs to frontiers of human
knowledge in combination of cosmology, astrophysics, and particle
physics, and can provide better understanding of the universe from
most small to most big, i.e., connecting quarks to the cosmos. The
universe is not empty, but full of background relic particles from
the big bang. It has long been anticipated that the highest energy
cosmic rays would be protons from outside the galaxy, and there is
an upper limit of the highest energy in the observed proton
spectrum, commonly referred to as the GZK cutoff \cite{GZK}, as the
protons traveling from intergalactic distances should experience
energy losses owing to pion productions by the photons in the cosmic
background radiation. Although there have been attentions for the
cosmic ray events above the GZK cutoff, it is natural to expect that
these ultrahigh energy cosmic rays come from sources within the GZK
zone \cite{Stecker}, i.e., not far from us in more than tens of Mpc.
Recently there are also reports on the observation of the GZK
cut-off by new experiments~\cite{HiRes}. However, questions about
the composition of such ultrahigh energy cosmic ray particles, e.g.,
whether they are protons, neutrons, or anti-nucleons~\cite{HM}, are
still open to investigations.

Muons in the air showers are mainly from decays of pions and kaons
produced in the interactions of the primary cosmic rays with the
atmosphere. The very high energy secondary pion and kaon cosmic rays
can be considered as from the current fragmentation of partons in
deep inelastic scattering of the primary cosmic rays with the
nucleon targets of the atmosphere in a first approximation
\cite{MSSY}. We also consider only the favored fragmentation
processes, i.e., the $\pi^+$, which is composed of valence $u$ and
$\bar{d}$ quarks, is from the fragmentation of $u$ and $\bar{d}$
quarks in the nucleon beam, and the $\pi^-$, which is composed of
valence $\bar{u}$ and $d$ quarks, is from the fragmentation of
$\bar{u}$ and $d$ quarks \cite{MSY}. Similarly, the $K^+$, which is
composed of valence $u$ and $\bar{s}$, is from the fragmentation of
$u$ and $\bar{s}$ quarks, and the $K^-$, which is composed of
valence $\bar{u}$ and $s$, is from the fragmentation of $\bar{u}$
and $s$ quarks. The $\mu^+$ is from the decay of a $\pi^+$ or a
$K^+$ and the $\mu^-$ is from the decay of a $\pi^-$ or a $K^-$. We
can roughly estimate the muon charge ratio by
\begin{equation}
\frac{\mu^+}{\mu^-} =\frac{\int_0^1 {\mathrm d} x \left\{\left[
u(x)+\bar{d}(x)\right]+\kappa
\left[u(x)+\bar{s}(x)\right]\right\}}{\int_0^1 {\mathrm d} x
\left\{\left[d(x)+\bar{u}(x)\right]+\kappa
\left[\bar{u}(x)+s(x)\right]\right\}}, \label{QME}
\end{equation}
where $q(x)$ is the quark distribution with flavor $q$ for the
incident hadron beam and $\kappa \sim 0.1 \to 0.3$ is a factor
reflecting the relative muon flux and fragmentation behavior of
$K/\pi$. Secondary collisions do not influence the above estimation,
since the current parton beams still keep their flavor content and
act as the current partons after the strong interactions with the
partons in the atmosphere targets. Adopting a simple model
estimation of the parton flavor content in the nucleon without any
parameter \cite{ZZM}, we find that $\mu^+/\mu^- \sim 1.7$ for proton
and $\mu^+/\mu^- \sim 0.7$ for neutron. This simple evaluation is in
agreement with the empirical expectation of $\mu^+/\mu^- \approx
1.66$ for proton and $\mu^+/\mu^- \approx 0.695$ for neutron
\cite{Adair} as well as that in an extensive Monte Carlo calculation
\cite{neutron-muon}, thus it provides a clear picture to understand
the dominant features for the muon charge ratio by the primary
hadronic cosmic rays. For the $\mu^+/\mu^-$ ratio for antiproton, it
is equivalent to the $\mu^-/\mu^+$ ratio for proton by using
Eq.~(\ref{QME}), thus we find $\mu^+/\mu^- \sim 0.6$ for antiproton,
which is close to that for neutron. The $\mu^+/\mu^-$ ratio for
antineutron is also equivalent to the $\mu^-/\mu^+$ ratio for
neutron, and it is $\mu^+/\mu^- \sim 1.4$, which is close to that
for proton. It is hard to distinguish between the primary neutrons
and antiprotons (or protons and antineutrons) by the $\mu^+/\mu^-$
ratio of the air shower, unless very high precision measurement is
performed and also our knowledge of the muon charge ratio for each
nucleon species is well established.

The study of cosmic rays with primary energies above $10^5$\,GeV are
typically based on the measurements of extensive air showers (EAS)
that they initiate in the atmosphere. The ground detector array
records the secondary particles produced in shower cascades,
including photons, electrons (positrons), muons, and some hadrons.
Then their arrival times and density profiles are used to infer the
primary energy and composition of the incident cosmic ray particle,
usually through comparison with simulated results. Photons,
electrons and positrons are the most numerous secondary particles in
an EAS event. However, for very inclined showers, these
electromagnetic components would travel a long slant distance and
are almost completely absorbed before they reach the ground. On the
other hand, muons are decay products of charged mesons in shower
hadronic cascades. Most high energy muons survive their propagation
through the slant atmospheric depth, during which they lose
typically a few tens of GeV's energy. These high energy muons carry
important information about the nature of the primary cosmic ray
hadron, which will be extracted from their energy spectrum and
lateral distribution.

As discussed by Hwang and I \cite{HM}, the ratio of positive versus
negative muons $\mu^+ / \mu^-$ is a significant quantity which can
help to discern the primary composition, and at high energies this
charge ratio also reflects important features of hadronic meson
production in cosmic ray collisions. In order to obtain such muon
charge information, we would need a way to distinguish between
positive and negative high energy muons. Unfortunately, existing
muon detectors available at shower arrays, usually scintillators and
water \v Cerenkov detectors, are not commonly equipped with
magnetized steel to differentiate the muon charges. Even if they
were, the limited region of the magnetic field prevents definite
determination of high energy muons' track curvature.

This invites us to think of the geomagnetic field as a huge natural
detector for muon charge information. Apparently, after being
produced high in the atmosphere, a positively charged muon would
bend east on its way down while a negatively charged muon would bend
west, introducing an asymmetry into the density profile of the
shower front. If their separation is large enough as compared with
other circularly symmetric ``background'' deviations, it will be
possible to distinguish the positive muons from the negative ones.

To see such an effect, Xue and I \cite{XM} analyzed the possibility
of obtaining the charge information of high energy muons in very
inclined extensive air showers. We have demonstrated that positive
and negative high energy muons in sufficiently inclined air showers
can be distinguished from each other through their opposite
geomagnetic deviations in the transverse plane. We developed a
revised Heitler model to calculate this distinct double-lobed
distribution, and studied the condition for the two lobes of either
positive or negative muons to be separable with confidence. From our
criterion of resolvability, we concluded that a zenith angle
$75^\circ \leq \theta \leq 85^\circ$ will be most suitable for our
approach.

There are already some results from full air shower simulations that
take into account the geomagnetic effect on muon propagation
\cite{Ave:2000xs,Aly:1964,Horandel:2003vu,Ayre:1972nx,Hebbeker:2001dn,Ostapchenko:2004qz,Apel:2005ch,Hansen:2004kf}.
They illustrated remarkable double-lobed muon lateral density
profile in very inclined air showers, which is in agreement with our
expectation qualitatively. However, no present study has fully
considered the high energy part of muon content, which can be used
to compare with our results. Thus we would like to propose future
simulations of very inclined extensive air showers that focus on the
behavior of high energy muons. They also have to keep track of the
muon charges and the relation to their lateral positions. For more
detailed analysis and discussion, please refer to Ref.\cite{XM}.

In summary, we propose to extract the charge information of high
energy muons in very inclined extensive air showers by analyzing
their relative lateral positions in the shower transverse plane.
This muon charge information is helpful to detect the composition of
cosmic rays, e.g., the neutron or antiproton content of the
ultrahigh energy cosmic rays.

\section*{Acknowledgments}

I am very grateful to Pauchy Hwang and the organizers for their
invitation and warm hospitality. I also thank Pauchy Hwang and
BingKan Xue for the collaborated results in this talk. This work is
partially supported by National Natural Science Foundation of China
(Nos.~10721063, 10575003, 10528510), by the Key Grant Project of
Chinese Ministry of Education (No.~305001), and by the Research Fund
for the Doctoral Program of Higher Education (China).


\begin{thebibliography}{0}


\bibitem{GZK}
K.~Greisen, Phys.\ Rev. \ Lett. {\bf 16}, 748 (1966); G.T.~Zatsepin
and V.A.~Kuzmin, Pis'ma Zh.\ Eksp.\ Tero.\ Fiz. {\bf 4}, 114 (1966)
[JETP Lett. {\bf 4}, 78 (1966)].

\bibitem{Stecker}
F.W.~Stecker, Phys.\ Rev.\ Lett. {\bf 21}, 1016 (1968). For an
extensive review, see, F.W.~Stecker, astro-ph/0101072.

\bibitem{HiRes}
R. Abbasi {\it et al.}, HiRes Collaboration, Phys.\ Rev.\ Lett. {\bf
100}, 101101 (2008). Also Pierre Auger Observatory reported the
observation of the GZK cut-off.

\bibitem{HM}
W-Y.P. Hwang and B.-Q. Ma,  Eur. Phys. J. A {\bf 25}, 467 (2005).

\bibitem{MSSY}
See, e.g., B.-Q.~Ma, I.~Schmidt, J.~Soffer, and J.-J.~Yang, Nucl.\
Phys.\ A {\bf 703}, 346 (2002).

\bibitem{MSY}
See, e.g., B.-Q.~Ma, I.~Schmidt, and J.-J.~Yang, Phys.\ Rev.\ D {\bf
65}, 034010 (2002).

\bibitem{ZZM}
Y.-J.~Zhang, B.~Zhang, and B.-Q.~Ma, Phys.\ Lett.\ B {\bf 523}, 260
(2001).

\bibitem{Adair}
R.K.~Adair, Phys.\ Rev.\ Lett. {\bf 33}, 115 (1974); R.K.~Adair {\it
et al.}, Phys.\ Rev.\ Lett. {\bf 39}, 112 (1977); O.C.~Allkofer {\it
et al.}, Phys.\ Rev.\ Lett. {\bf 41}, 832 (1978).


\bibitem{neutron-muon}
J.N.~Capdevielle and Y.~Muraki, Astropart.\ Phys. {\bf 11}, 335
(1999). See, e.g., Fig.~7.

\bibitem{XM}
B. Xue and B.-Q. Ma, Astropart. Phys. {\bf 27}, 286 (2007).





\bibitem{Ave:2000xs}
  M.~Ave, R.~A.~Vazquez and E.~Zas,
  %``Modelling horizontal air showers induced by cosmic rays,''
  Astropart.\ Phys.\  {\bf 14}, 91 (2000).
%  [arXiv:astro-ph/0011490].
  %%CITATION = ASTRO-PH 0011490;%%

\bibitem{Aly:1964}
  H.~H.~Aly, M.~F.~Kaplon, and M.~L.~Shen,
  Nuovo.\ Cim.\ {\bf 31}, 905 (1964).

\bibitem{Horandel:2003vu}
  J.~R.~Horandel,
  %``On total inelastic cross-sections and the average depth of the maximum
  %of extensive air showers,''
  J.\ Phys.\ G {\bf 29}, 2439 (2003).
%  [arXiv:astro-ph/0309010].
  %%CITATION = ASTRO-PH 0309010;%%

\bibitem{Ayre:1972nx}
  C.~A.~Ayre {\it et al.},
  %``Multiple scattering of cosmic ray muons in the range 10-70 gev/c,''
  J.\ Phys.\ A {\bf 5}, L102 (1972).
  %%CITATION = JPAGB,A5,L102;%%

\bibitem{Hebbeker:2001dn}
  T.~Hebbeker and C.~Timmermans,
  %``A compilation of high energy atmospheric muon data at sea level,''
  Astropart.\ Phys.\  {\bf 18}, 107 (2002).
%  [arXiv:hep-ph/0102042].
  %%CITATION = HEP-PH 0102042;%%

\bibitem{Ostapchenko:2004qz}
  S.~Ostapchenko,
  %``QGSJET-II: Results for extensive air showers,''
  Nucl.\ Phys.\ Proc.\ Suppl.\  {\bf 151}, 147 (2006).
%  [arXiv:astro-ph/0412591].
  %%CITATION = ASTRO-PH 0412591;%%

\bibitem{Apel:2005ch}
See, e.g.,
  W.~D.~Apel {\it et al.}  [KASCADE Collaboration],
  %``Comparison of measured and simulated lateral distributions for  electrons
  %and muons with KASCADE,''
  arXiv:astro-ph/0510810.
  %%CITATION = ASTRO-PH 0510810;%%

\bibitem{Hansen:2004kf}
  P.~Hansen, T.~K.~Gaisser, T.~Stanev and S.~J.~Sciutto,
  %``The influence of the geomagnetic field and of the uncertainties in the
  %primary spectrum on the development of the muon flux in the atmosphere,''
  Phys.\ Rev.\ D {\bf 71}, 083012 (2005);
%  [arXiv:astro-ph/0411634];
  %%CITATION = ASTRO-PH 0411634;%%
  A.~Cillis and S.~J.~Sciutto,
  %``Geomagnetic field and air shower simulations,''
  arXiv:astro-ph/9908002.
  %%CITATION = ASTRO-PH 9908002;%%


\end{thebibliography}
\end{document}